\documentclass{article}

\usepackage{microtype}
\usepackage{graphicx}
\usepackage{subfigure}
\usepackage{booktabs} 

\usepackage{hyperref}

\usepackage[accepted]{icml2018}

\icmltitlerunning{Harmonic Recomposition using Conditional Autoregressive Modeling}

\begin{document}

\twocolumn[
\icmltitle{Harmonic Recomposition using Conditional Autoregressive Modeling}




\begin{icmlauthorlist}
\icmlauthor{Kyle Kastner}{udem}
\icmlauthor{Rithesh Kumar}{udem}
\icmlauthor{Tim Cooijmans}{udem}
\icmlauthor{Aaron Courville}{udem,cifar}
\end{icmlauthorlist}

\icmlaffiliation{udem}{Montr\'eal Institute for Learning Algorithms (MILA), Universit\'e de Montr\'eal, Montr\'eal, Qu\'ebec, Canada}
\icmlaffiliation{cifar}{CIFAR Fellow}

\icmlcorrespondingauthor{Kyle Kastner}{kyle.kastner@umontreal.ca}

\icmlkeywords{Machine Learning, ICML}

\vskip 0.3in
]



\printAffiliationsAndNotice{}  

\begin{abstract}
We demonstrate a conditional autoregressive pipeline for efficient music recomposition, based on methods presented in \citet{vandenoordvqvae2017}. Recomposition  \citep{casalrecomposition2010} focuses on reworking existing musical pieces, adhering to structure at a high level while also re-imagining other aspects of the work. This can involve reuse of pre-existing themes or parts of the original piece, while also requiring the flexibility to generate new content at different levels of granularity. Applying the aforementioned modeling pipeline to recomposition, we show diverse and structured generation conditioned on chord sequence annotations. 
\end{abstract}

\section{Introduction}
Since the early days of computation, composers have explored methods of combining aleatoric music and algorithmic composition with generic computing devices \citep{xenakisformal2003,copeemi1989}. 
Authors have taken a wide variety of data driven approaches to "creative generation" in various domains \cite{barbierilyrics2012,ha2017neural,Graves2013}, with extensive application to music modeling \cite{briotsurvey2017,robertsmusicvae2018,eck+schmidhuber:icann2002,sturm15ismir,hadjeres2016deepbach,boulanger2012modeling,bretandialogue2017,robertsmusicvae2018}.

In this paper, we focus on the task of \emph{harmonic recomposition} \citep{casalrecomposition2010}. Melody generation and evaluation is a difficult task, even in monophonic music \cite{jaques2016sequence}, so we use the term
\emph{harmonic} recomposition to reference our focus on aspects of agreement and structure between voices.  
Our pipeline is also applicable to purely sequential and iterative generation, as has been shown in prior work \citep{huangcoconet2017,vandenoordvqvae2017}.

\subsection{Related Work}
Autoregressive models have proven to be powerful distribution estimators for images and sequence data, showing excellent results in generative settings \citep{wavenet}. They have also performed well in related prior work for polyphonic music generation \cite{briotsurvey2017}.
Most related to the work described in this paper is CoCoNet \citep{huangcoconet2017, huang2018towards}, which also uses an autoregressive convolutional model over image-like structures for polyphonic music generation and was a direct inspiration for our approach. One key difference of our approach is our utilization of a two stage pipeline (first seen in the work of \citet{vandenoordvqvae2017}) which greatly improves training and generation speed as well as creating an implicit separation between local voice agreement (first stage) and global consistency over measures (second stage). 
\section{Implementation Details}
In this section, we describe the data, model, and training details for our recomposition approach.  An open source implementation of our setup (including audio samples) is available online\footnote{\url{https://github.com/kastnerkyle/harmonic_recomposition_workshop}}.
\subsection{Data}
We use a subset of the scores associated with the composer Josquin des Prez as compiled by the Josquin project \footnote{\url{http://josquin.stanford.edu/}}.
Only pieces with $4$ parts are considered, resulting in a dataset with $103$ pieces comprising $5568$ measures. We hold a contiguous $500$ measures out for use as a source of harmonic chord sequences during conditional generation.

\begin{figure}[ht]
\centering
\begin{minipage}[b]{0.4\linewidth}
\centering
\includegraphics[height=.9\textwidth,width=\textwidth]{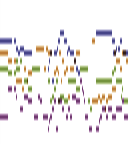}
\end{minipage}
\hspace{0.4cm}
\begin{minipage}[b]{0.4\linewidth}
\centering
\includegraphics[height=.9\textwidth,width=\textwidth]{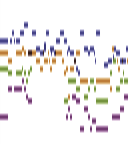}
\end{minipage}
\caption{Two 8 bar sequences generated over the conditioning sequence $(III)$,$III$, $v$, $iv$, $iv$, $v$, $i$, $iv$, $III$, $(III)$ using different random seeds}
\label{fig:figure1}
\end{figure}

After extracting individual measures, we convert to a "piano roll" style multichannel image, with each measure having $16$ quantized timesteps (regardless of time-signature) on the horizontal axis, and one of $49$ possible tones on the vertical axis, where $49$ comes from the set of all possible notes used in the key normalized data \citep{hadjeres2016deepbach}. These $49$ values are padded to $52$ for compatibility with the convolutional strided layers used in the VQ-VAE, and each voice assigned its own channel in an image-like container of size $(N, 52, 16, 4)$ described in examples, height, width, and channels format (NHWC). The overall result can be seen in Fig. \ref{fig:figure1}, where each color represents a separate channel.


\subsection{Conditional Information}
We extract the chord function and voicing of all measures using the music21 software package \citep{cuthbertmusic21}, and form "function triplets" of the previous, current, and next measure.
A measure group of $I, ii^4_2, V^6_5$ chords would form triplet groupings of $I, I, ii^4_2$ (we repeat chords to handle border issues), then $I, ii^4_2, V^6_5$, and finally $ii^4_2, V^6_5, V^6_5$. 

\subsection{Models}
The model pipeline is a two-stage generative setup, as described by \citet{vandenoordvqvae2017}, wherein an initial stage (denoted VQ-VAE) is unconditionally trained to compress inputs to a spatially reduced, discrete representation (which we call $Z$), and uncompress. Once the VQ-VAE stage is trained, we use it to generate a compressed $Z$ for each element in the dataset, and train an autoregressive generative "prior model" on this representation. The prior model learns to generate components of $Z$ (which takes the form of a $(13, 4)$ spatial map in this work), denoted $Z_{i, j}$, one at a time conditioning the next generation step on all previously generated $Z_{<i,<j}$. 

\begin{figure}[ht]
\centering
\begin{minipage}[b]{0.4\linewidth}
\centering
\includegraphics[height=.9\textwidth,width=\textwidth]{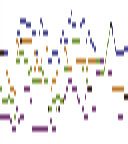}
\end{minipage}
\hspace{0.5cm}
\begin{minipage}[b]{0.4\linewidth}
\centering
\includegraphics[height=.9\textwidth,width=\textwidth]{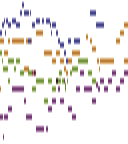}
\end{minipage}
\caption{8 bar sequences generated over conditioning sequence 
$(ii)$,$I$,$V$,$vi$,$ii$,$V$,$ii$,$ii$,$i5$,$(I)$. \emph{Left}: only chord triplet conditioning. \emph{Right}: chord triplet conditioning along with previous generation as conditioning spatial map.}
\label{fig:figure2}
\end{figure}

The prior model may also take conditioning as one or multiple vectors (separate embeddings for each previous, current, and next chord, indexed by a chord integer), a spatial map (a $Z$ from some previous measure), or a combination of both during the generation process. The effect of conditioning type can be seen in Fig \ref{fig:figure2}.


\subsection{Experiment Details}
The first stage VQ-VAE has 2 strided convolutional layers of kernel size $(4, 4)$ and a stride of $2$ on both spatial axes, followed by an additional 2 layers of $(4, 4)$ convolution with stride $1$, using rectified linear activations \citep{glorotrelu2011} and batch normalization \citep{ioffebn2015}. These layers have sizes $64$, $128$, $257$, and $256$ leading to a VQ codebook size of $256$, which results in a latent $Z$ dimension of $(13, 4)$ for size $(52, 16, 4)$ input. This procedure is inverted using transpose convolution for the decoder and combined with a binary cross-entropy loss alongside codebook and commitment losses for the VQ-VAE, averaged over all channels and spatial dimensions. Training was performed over $50000$ minibatches of size $64$ with an Adam optimizer with $\alpha=0.0002$, $\beta1=0.9$, $\beta2=0.999$, $\epsilon=1E-8$ \citep{kingma2014adam}.  

In the second stage, a gated conditional PixelCNN \citep{vandenoordpixelcnn2016} is configured with $15$ layers of $64$ projection channels. The first layer has a kernel size of $(5, 5)$ and no residual connection. Layers after the first utilize residual connections \citep{heres2016} and have a kernel size of $(3, 3)$, followed by $(1, 1)$ convolution, rectified linear activation, and a final convolution of size $(1, 1)$ and $256$ channels (due to the VQ codebook size of $256$). Training was performed over $100000$ minibatches of size $50$ with an Adam optimizer (configured as before) with categorical cross-entropy loss averaged over the output.

\section{Results}
We experiment with two types of conditioning combined with the aforementioned architecture. The higher level information contained in the chord sequences alone seems sufficient to produce directed, coherent trajectories, without need for spatial conditioning information. When spatial conditioning from the previous timestep is included, the resulting generations are punctuated by dissonant intervals or long silent gaps. Finding better ways to combine local note level information with chord annotation will be an important step to improving this pipeline.

\vspace{-0.37cm}
\vspace{-.5cm}
\vspace{-0.5cm}

\section{Conclusion}
Chord-conditional generative models are an ideal fit for harmonic recomposition. We find that a two-stage pipeline reminiscent of \citet{vandenoordvqvae2017} and \citet{huangcoconet2017} captures musical structure, and allows for chord conditional generation. Our work demonstrates note-level realizations of given chordal sequences and provides an open-source implementation with examples. 


\newpage
\nocite{raffel2015large}

\bibliography{bib}
\bibliographystyle{icml2018}
\end{document}